# Material surface – analyte interactions with similar energy rates vary as univariate quadratic function of topological polar surface area of analytes

Nirman Chakraborty, Sagnik Das, Debdulal Saha and Swastik Mondal*

CSIR Central Glass and Ceramic Research Institute, 196, Raja S. C. Mullick Road, Kolkata-700032, India

*Corresponding author: swastik_mondal@cgcri.res.in

**Abstract**

**Material surface - analyte interactions play important roles in numerous processes including gas sensing. However, the effects of topological polar surface area (TPSA) of analytes on surface interactions during gas sensing have been so far largely disregarded. In this work, based on experimental observations on changes in electrical resistance of cadmium sulphide (CdS) due to surface interactions during gas sensing, we found that unexpected univariate quadratic correlation exists between changes in resistance of CdS and TPSA of analytes. Further experiments on four other material systems showed the same trend, revealing a generalized picture of TPSA dependence of surface interactions.**

The key to several material functionalities lie with the interaction of the surface with target analytes [1-3]. Be it gas sensing [4-5], permeability of drugs [6], absorption through membranes [7] or water purification [8]; the primary interplay at surface plays a pivotal role in deciding the route of reaction that eventually leads to material performance [9-14]. Despite the importance, the mechanisms of interaction between surfaces and the reacting chemical species (analytes) are poorly understood. Different surface phenomenon proceed via different routes; like catalysis and absorption through membranes proceed via charge exchange with surface which eventually leads to generation of reaction by-products that get detected either by spectroscopic methods like UV-Vis, NMR, chromatography or by electrochemical techniques [15-17]. On the other hand, in mechanisms like chemiresistive or capacitive gas sensing, the extent of change in electrical property of material on surface interaction with target gas determines how sensitive the material is.

The electrical resistance of a thin or thick film material dynamically changes when its surface undergoes chemical interactions with analytes (see Fig. 1, Supplementary Information). A greater difference in electrical resistance in presence of an analyte and without it indicates a greater surface interaction. This phenomenon is most commonly used for chemiresistive gas sensing, where the gas sensing response is defined based on change in dynamic resistance of materials due to surface reactions as S = $\frac{R_a - R_g}{R_a}$ (for surface interaction with reducing gases) or S $=\frac{R_a - R_g}{R_g}$ (for surface interaction with oxidising gases), where $R_a$ and $R_g$ are resistances of the sensing material in presence of air and target analyte respectively [18-19]. So far, improved surface interactions for gas sensing have been achieved empirically by modifying morphology, composition, temperature, particle size, pore size and pore distribution in material surfaces [4,12,18]. However, the effects of target analytes, particularly contributions from polar atoms in analytes on surface interactions during gas sensing have been largely disregarded. As evident from the above formulae, the change in resistance is directly proportional to the sensing response. Since the surface reactions occur due to effective interaction of analyte molecules with chemisorbed surface oxygen layer, smaller polar area of target gas should ensure greater interaction in terms of number of surface attachments. This possibility prompts us to use the concept of TPSA (see Supplementary Information, Table I) of gas molecule [20,21] in different gas sensing phenomenon; which is based on the summation of surface contributions by polar atoms in a molecule and has been predominantly used in predicting molecular transport of drugs through different body membranes (see Supplementary Information). Hence, based on TPSA concept, number of molecules with lower TPSA should have more possibility of surface adsorption compared to the higher ones [22,23]. In order to delineate the role of TPSA on surface interactions, we performed dynamic resistance measurements on thick films of 4 pristine and 1 doped materials for surface interactions with 17 different analytes having various TPSA values ranging from 1 to 40.5 Å$^2$ (see Fig.1, 2, 3, Supplementary Information).

Since material surface - analyte interactions depends on several factors, it is necessary to have a common ground based on which a comparison of the effects of TPSA on surface interactions can be made. In the present work, we have used similar energy rates of surface interactions as the criterion for such comparison. For energy rates calculations, the commonly used Arrhenius equation for activation energy has been employed [24]

$$R = R_0 e^{\frac{E}{KT}} \quad \ldots\ldots\ldots\ldots\ldots (1)$$

where R is the material resistance in Ω, $R_0$ is the intrinsic material resistance that depends on bulk material properties, E is the energy in eV, K is Boltzmann constant and T is the temperature in Kelvin; we have considered the sample resistance both in presence and absence of target gas. When the material is under normal conditions, eq. (1) becomes

$$R_{air} = R_0 e^{\frac{E_{air}}{KT}} \quad \ldots\ldots\ldots\ldots (2),$$

where $R_{air}$ is the resistance of the sensor material in air and the surface energy of material is $E_{air}$. In the presence of target gas, the equation becomes

$$R_{gas} = R_0 e^{\frac{E_{gas}}{KT}} \quad \ldots\ldots\ldots\ldots (3),$$

where the resistance of the sensor material is $R_{gas}$ and surface energy of material is $E_{gas}$. Dividing (2) by (3) gives

$$\frac{R_{air}}{R_{gas}} = e^{\frac{\Delta E}{KT}} \quad \ldots\ldots\ldots\ldots\ldots\ldots (4),$$

where $\Delta E = E_{air} - E_{gas}$. Hence, it is this change in energy involved with the surface reaction that is actually a manifestation of the ease of occurrence of the reaction in terms of lowering the activation energy. For gas detection by chemiresistive methods in particular, response time is an important factor that indicates how fast the reaction occurs and it is manifested as rate of change in sample resistance. Thus a scaling of the above change in energy in terms of response time provides the actual energy rate of surface reaction that we termed as ERSI (see Table I, Supplementary information).

A detailed description of all the samples chosen along with sensor fabrication strategy and other characterization tools used are provided in Supplementary Information. The dynamic response curve (resistance vs. time graph) for each gas/VOC sensing was used to obtain the values of $R_{air}$ (sensor resistance in absence of gas) and $R_{gas}$ (sensor resistance in presence of gas) along with the τ (response time, calculated in terms of time period required by the sensor to reach minimum/maximum resistance on application of target gas/VOC). Then using formula 4, the energy change involved is obtained as $\Delta E = K \times T \times \ln(\frac{R_{air}}{R_{gas}})$. ERSI is finally obtained by dividing $\Delta E$ with τ (see Supplementary Information). TPSA values of different molecules have been obtained from PubChem [21] database, where the values were calculated using Cactvs and OpenEye softwares. The estimations were based on summation of tabulated surface contributions of polar fragments in the target molecule. The contributions of individual polar fragments were fitted by least squares method and compared with 3D PSA values (see Table I, Supplementary information) [20, 21].

Based on the above concepts of ERSI and TPSA, the sensing responses towards different gases by various materials can now be discussed (see Table I). For the cadmium sulphide sample, calculations based on ERSI for all the gases reveal that for ammonia, pyran, formaldehyde, NO, iso-amyl alcohol and $SO_2$, the ERSI are similar in range of 0.001-0.003 eV/S. Responses initially increase as we go from ammonia to pyran with maximum 24.3% and then follows a decreasing trend (see Fig. 2, Table I, Supplementary information). In present work, we have chosen the lowest TPSA limit to 1 Å$^2$ as per the database [23] considering it as a minimum requirement for effective interaction with polar surface oxygen layer. While ammonia has a low TPSA of 1 Å$^2$, followed by 9.2 for pyran, 17.1 for formaldehyde, 18.1 for nitric oxide, 20.2 for iso-amyl alcohol and 35 for $SO_2$, the gases with similar ERSI show responses depending on their TPSA; the molecules with high TPSA have lower response, however the vice-versa is not true (see Fig. 2, Table I, Supplementary information). For certain range, the response increases with increasing TPSA and starts decreasing with further increase in TPSA values (see Fig. 2, Supplementary information). Looking from the other way round, carbon monoxide, hydrogen sulphide, ammonium chloride and ammonia have the same TPSA of 1 Å$^2$. On comparing their ERSI with sensing responses, we can observe a different correlation; ammonia with greater reaction rate has greater response. The similar correlation can also be found among benzophenone, formaldehyde, benzaldehyde and acetone which have similar TPSA of 17.1 Å$^2$ and acetone shows maximum response due to higher ERSI. For ethanol, methanol, iso-amyl alcohol and iso-propanol which have the same TPSA of 20.2 Å$^2$, maximum response is exhibited by ethanol. Thus for target analytes with similar ERSI, the sensing response follow an unexpected correlation with TPSA. And for molecules with similar TPSA, the sensing response is somewhat proportional to ERSI (see Fig. 3, Table I, Supplementary information).

In order to test the general applicability of our correlation in other material systems, we performed further experiments at different operating temperatures using the same set of 17 analytes on four different materials $SnO_2$, $Sn_{0.696}V_{0.304}O_2$ [25], $TiO_2$ and AlN (Table I, Supplementary information) having different structures, particle size and morphologies (see Supplementary information). For pure $SnO_2$, gas sensing experiments at 350°C with 10 ppm of different gases revealed that its response was maximum for ammonia with 33.5% response with cross-responses to several gases (see Supplementary information). While for analytes like $H_2S$, pyran, benzaldehyde, NO, Iso-amyl alcohol and $SO_2$ the ERSI during response is in the range of 0.002-0.004 eV/S, the correlation between sensing response and TPSA differs from expected inverse relationship [15] particularly for low TPSA values; $H_2S$ with lowest TPSA of 1 Å$^2$ has response of 13% and a maximum response of 14% has been observed for pyran with TPSA 9.2 Å$^2$ (see Fig. 2). For benzaldehyde, NO, Iso-amyl alcohol and $SO_2$, the response decreases with increasing TPSA. In case of analytes with similar TPSA (CO, $H_2S$, $NH_4Cl$ and ammonia with TPSA of 1 Å$^2$; benzophenone, formaldehyde, benzaldehyde and acetone with TPSA 17.1 Å$^2$; methanol, iso-amyl alcohol, iso-propyl alcohol and ethanol with TPSA 20.2 Å$^2$), the sensing responses are somewhat proportional to ERSI values (Table I, see Supplementary information). The similar correlation was also observed for pure $Sn_{0.696}V_{0.304}O_2$, $TiO_2$ in anatase phase and pure AlN, establishing the generality of the correlation to other material systems (Table I, see Supplementary information). However, in the doped variant $Sn_{0.696}V_{0.304}O_2$, the selectivity could be increased indicating the importance of doping in customizing materials for improving a particular function.

In order to identify the nature of relationship that ERSI and TPSA hold with sensing response, data points obtained from above experiments were fitted by mathematical functions (see Fig. 2, 3, Supplementary information). By fitting the curves of material response vs. ERSI for particular TPSA, it has been observed that response is proportional to half power of ERSI (Fig. 3,

Supplementary information). And the curves obtained by fitting material response vs. TPSA values for a particular ERSI range reveal that the change in response with TPSA follows a univariate quadratic polynomial function [26] (see Fig. 2, Supplementary information). This quadratic correlation of material surface – analyte interactions with TPSA is remarkable as it implies that for certain range of TPSA, the response or magnitude of surface interaction would increase with increasing TPSA, which is in contrast to the general understanding that surface interaction should decrease with increasing TPSA [27]. Such understanding can be used, for example, to modify the TPSA of a target analyte by changing the structures of non-essential chemical groups in the analyte to achieve improved responses for a given surface [28]. Furthermore, since TPSA is an intrinsic property of a chemical species and the quadratic relationship holds true irrespective of different surface topography, composition, structure, particle size and morphology of the material (Supplementary information), it is expected that similar correlation should be applicable to other processes where surface interaction is involved, such as, absorption of molecules through a surface, catalysis or permeation of molecules through a membrane. For example, this correlation might explain increased partition coefficient of drug molecules (depicting membrane permeability of a drug molecule) with increasing TPSA for small values of TPSA [23, 29]. Partition coefficients of drug molecules like amitriptyline (TPSA = 3.2 Å$^2$, log P = 5.04), imipramine (6.5 Å$^2$, log P= 4.42), methamphetamine (12 Å$^2$, log P= 2.1), nicotine (TPSA = 16.1 Å$^2$, log P= 1.2) in comparison to ammonium chloride (TPSA = 1 Å$^2$, log P= -0.21) follow a similar trend as discovered in this work. Also in case of membrane separation where partition coefficient of target gases [23] play an important role [30], the gases/VOCs like ammonia (TPSA = 1 Å$^2$, log P=-0.7), pyran (TPSA = 9.2 Å$^2$, log P=1.2), acetone (TPSA = 17.1 Å$^2$, log P=-0.1) and ethylene glycol (TPSA = 40.5 Å$^2$, log P=-1.4) show the same trend. Our results thus establish that for physicochemical processes that involve materials surface, parabolic correlation between surface interaction and polar surface area of analytes needs to be considered. Furthermore, the dynamic polar surface area [27, 31] or experimental polar surface area obtained from X-ray charge density analysis [32, 33] and other affecting parameters besides energy rates can be examined to get more refined models of surface interactions.

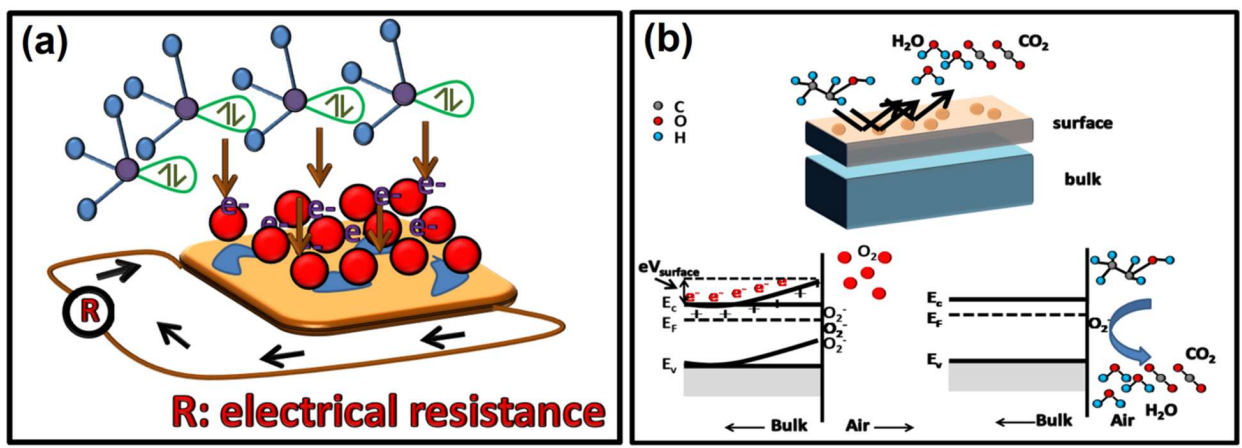

FIG. 1. Chemiresistive gas sensing mechanism. (a) A pictorial representation of electrical network to probe surface interactions. (b) Mechanism of band bending explaining the cause of sample resistance due to interaction at surface.

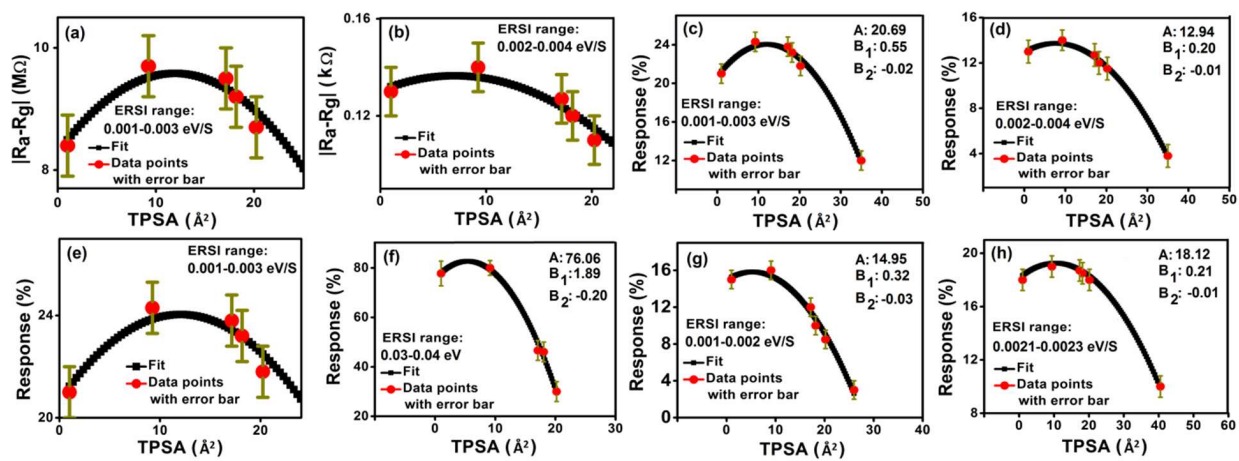

FIG. 2. Quadratic ($y = Ax + B_1 x + B_2 x^2$, y=response, x=TPSA) correlation between surface interaction and TPSA: range in which response increases with TPSA and then decreases. Change in resistance vs. TPSA (up to 20 Å$^2$) plots for (a) CdS (b) SnO$_2$; response vs. TPSA (up to 20 Å$^2$) plot for (e) CdS. Quadratic ($y = Ax + B_1 x + B_2 x^2$, y=response, x=TPSA) correlation between surface interaction and TPSA: full range. Response vs. TPSA plots for (c) CdS (d) SnO$_2$ (f) Sn$_{0.696}$V$_{0.304}$O$_2$ (g) TiO$_2$ (anatase) (h) AlN samples respectively.

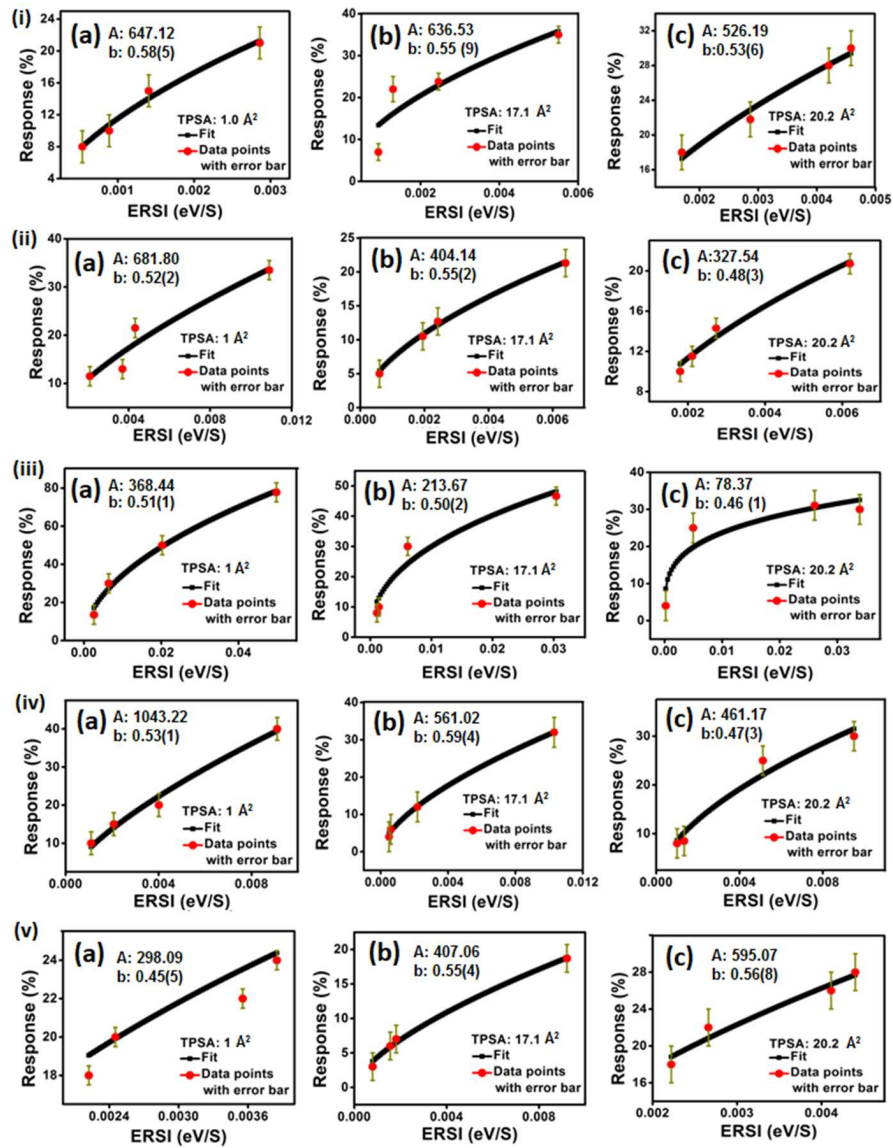

FIG. 3 (i) Response vs. ERSI curves for surface interactions with gas molecules of TPSA (a) 1 Å$^2$ (b) 17.1 Å$^2$ (c) 20.2 Å$^2$ for CdS(ii) SnO$_2$ (iii) Sn$_{0.696}$V$_{0.304}$O$_2$ (iv) TiO$_2$ (v) AlN. The fits are put in black colour and experimental data points in red. Uncertainty/error in data have been calculated by standard error propagation method and normalized with respect to maximum error. Error bars are presented in green.

TABLE I. Responses to gas/VOC with particular TPSA for interactions lying in a particular ERSI range. 9.2 Å$^2$ TPSA vapor for Sn$_{0.696}$V$_{0.304}$O$_2$ was THF and for rest it was pyran. This is actually a manifestation of selectivity by virtue of vanadium doping. All resistance values are in MΩ except for SnO$_2$ which is in kΩ.

| Material | ERSI range (eV/S) | TPSA (Å$^2$) | Gas/VOC | \|R$_a$-R$_g$\| | Response (%) |
|---|---|---|---|---|---|
| CdS | 0.001-0.003 | 1 | NH$_3$ | 8.4 | 21 |
| | | 9.2 | Pyran | 9.7 | 24.3 |
| | | 17.1 | Formaldehyde | 9.5 | 23.8 |
| | | 18.1 | NO | 9.2 | 23.2 |
| | | 20.2 | Iso-amyl alcohol | 8.7 | 21.8 |
| | | 35 | SO$_2$ | 4.8 | 12 |
| SnO$_2$ | 0.002-0.004 | 1 | H$_2$S | 0.13 | 13 |
| | | 9.2 | Pyran | 0.14 | 14 |
| | | 17.1 | Benzaldehyde | 0.127 | 12.7 |
| | | 18.1 | NO | 0.12 | 12 |
| | | 20.2 | CO | 0.11 | 11.5 |
| | | 35 | SO$_2$ | 0.04 | 3.8 |
| Sn$_{0.696}$V$_{0.304}$O$_2$ | 0.03-0.04 | 1 | NH$_3$ | 6.24 | 77.77 |
| | | 9.2 | THF | 6.4 | 80 |
| | | 17.1 | Acetone | 3.76 | 46.62 |
| | | 18.1 | NO | 3.68 | 46.02 |
| | | 20.2 | Iso-propyl alcohol | 1.6 | 30 |
| TiO$_2$ | 0.001-0.002 | 1 | CO | 7.5 | 15 |
| | | 9.2 | Pyran | 8 | 16 |
| | | 17.1 | Formaldehyde | 6 | 12 |
| | | 18.1 | NO | 5 | 10 |
| | | 20.2 | Iso-amyl alcohol | 4.25 | 8.5 |
| | | 26 | Aniline | 1.5 | 3 |
| AlN | 0.0021-0.0023 | 1 | CO | 5.4 | 18 |
| | | 9.2 | Pyran | 5.7 | 19 |
| | | 17.1 | Acetone | 5.6 | 18.7 |
| | | 18.1 | NO | 5.5 | 18.5 |
| | | 20.2 | Methanol | 5.4 | 18 |
| | | 40.5 | Ethylene glycol | 3 | 10 |

**Acknowledgements** Authors are thankful to the Materials Characterization division, CSIR-CGCRI, for the XRD, FESEM analyses; Advanced Materials Characterization unit for TEM experiments; Central Materials Characterization division for the FTIR, TGA studies and Specialty Glass division for the Raman experiments. Authors acknowledge Md. Jalaluddin Mondal and Raju Manna for fabrication of sensor module and help during sensing experiments. N.C. acknowledges DST, Govt. Of India for INSPIRE fellowship (IF170810).S. M. acknowledges funding from SERB Core Research Grant, Govt. Of India (Grant number: CRG/2019/004588) and CSIR Mission mode project on "Food and Consumer Safety Solutions" (HCP0016).

**Author Contributions** S.M. conceived and supervised the research project. N.C. carried out all experiments. S.D. and D.S. developed the sensor substrates. N.C. and S.M. analyzed the experimental results. N.C. and S.M. wrote the manuscript with input from all authors.

**Competing interests** The authors declare no competing interests.

# Supplementary Information

**Material surface – analyte interactions with similar energy rates vary as univariate quadratic function of topological polar surface area of analytes**

Nirman Chakraborty, Sagnik Das, Debdulal Saha and Swastik Mondal[*]

*CSIR Central Glass and Ceramic Research Institute, 196, Raja S. C. Mullick Road, Jadavpur, Kolkata 700032, India*

*Corresponding author: swastik_mondal@cgcri.res.in

## Methods

### Sensor fabrication and sensing experiments

For preparing the sensor module, a small amount of sample (0.5 g) was made into consistent slurry in isopropyl alcohol and coated on a cylindrical substrate of 3 mm diameter and 4 mm length. The electrical contacts were made by platinum electrodes attached to the substrate by gold paste. A Nichrome wire was coiled inside the cylindrical substrate for heating. The substrate was then cured at 60°C for 6 hours and mounted on a 6 headed pin module, capped with a 200 μm polymeric mesh. The current-voltage responses were studied using an Agilent Multimeter interfaced with the software Agilent GUI data logger (see Fig. S1-S3). For vanadium doped tin oxide sensor, stable base resistance of the order of 8 MΩ was observed at 5V (350°C) (see Fig. S4) and hence the sensor was aged for 48 hours at 5V current supply. Since the vanadium doped sample showed maximum response to 10 ppm ammonia in air at 350°C, sensing experiments were performed at the same temperature with same concentration of target gas. The sensing experiments were performed in a fume hood placed inside the sensing chamber. For all the gases, standard calibrated cylinders were used, followed by a mass flow controller (Alicat) and nitrogen mixing chamber (see Fig. S2). Dynamic sensing measurements were carried out by applying 3 consecutive pulses of each gas at regular intervals, followed by regaining of steady electrical state. In order to check the repeatability of responses, the experiments were performed repeatedly for 120 days, at regular interval of 30 days (see Fig. S5). For avoiding errors due of carrier medium, blank experiments with air and nitrogen were performed and necessary corrections were made. Similar steps were also followed for pure $SnO_2$ and $TiO_2$ sensors, which were tested for 10 ppm of different gases using an operating temperature of 350°C. For CdS sample, stable base resistance was achieved at room temperature itself. In order to check the temperature independence of our correlation, similar studies on sensing responses were performed with CdS as well. For AlN, stable base resistance was observed at 250°C operational temperature and hence sensing experiments were performed at that temperature itself (see Fig. S4).

### Synthesis of sensor materials

Pure and vanadium doped tin oxide samples were prepared by a simple co-precipitation method followed by ultrasonication. In this method, measured amount of tin chloride ($SnCl_2 \cdot 2H_2O$, 99%, Merck) was dissolved in distilled water followed by adding few drops of conc. $HNO_3$ (Merck). Measured amount of vanadium pentaoxide ($V_2O_5$, Sigma Aldrich) was separately dissolved in distilled water followed by few drops of conc. $HNO_3$. The solution was allowed to undergo heating at 80°C till a clear solution was formed. The vanadium solution was added to tin solution followed by adding ammonia ($NH_4OH$, 25%, Merck) drop-wise till a pH of 9 was achieved. The solution was then ultrasonicated for an hour using ChromTech probe sonicator of power 360 W followed by calcination at 650°C. For TiO$_2$ (anatase) sample, 0.25 ml of titanium iso-propoxide (*Ti*[OCH(CH$_3$)$_2$]$_4$, Sigma Aldrich) was added to 20 ml of ethanol (Merck) followed by 10 ml of distilled water. The whitish precipitate formed was ultrasonicated for one hour followed by calcination at 450°C. The calcination temperature was determined by mass loss studies from room temperature to 880°C at the rate of 5°C/min. TGA analysis of all samples were carried out in a NETZSCH STA 449 C TGA analyzer. For analysis, small amount of sample (around 0.01 g) was taken and loaded in Al$_2$O$_3$ crucible. An empty alumina crucible was used as reference. The respective mass losses at different temperatures were identified by the boiling point of solvents/chemicals used in the synthesis. CdS and AlN samples were directly procured from Sigma Aldrich (99.9% pure). Presence of any impurity was checked by TGA analysis where negligible mass loss was observed (see Fig. S6) and it was attributed to the loss to moisture adsorbed by the samples due to their hygroscopic behaviour.

**Characterization**

X-ray diffractogram of all powder samples were collected using a X'PertPro MPD (PANalytical) diffractometer (θ-2θ mode, 2θ range for tin oxide samples was 10° to 110°, for TiO$_2$ the range was from 20° to 80°, CdS in range 10° to 100° and AlN in range 20° to 110°) at room temperature (see Fig. S7). For sample preparation, small amount of powder (1 g) was finely ground in an agate mortar and dried at 100°C to eliminate moisture. The powder samples were then mounted on a zero background holder and data was collected using Cu K$_α$ radiation. Lattice constants for all materials were obtained by Le' Bail[r1] refinement of the powder XRD data using Jana 2006 software[r2]. Morphological and microstructural analyses (see Fig. S8-S14) were performed in a TecnaiG2 TEM (Transmission Electron Microscopy) machine. For sample preparation, small amount of powder sample (0.001 g) was dispersed thoroughly in isopropyl alcohol solution and drop-casted on a copper coated carbon grid using a 10 μl pipette. The grids were allowed to dry overnight under normal conditions and then loaded in the TEM machine. For SnO$_2$ based samples, bright field images reveal formation of spherical nano-particles in range of 20-30 nm. High Resolution TEM (HRTEM) images reveal preferred exposure of 110 planes (see Fig. S9). For TiO$_2$ sample, bright field images reveal formation of spherical nanoparticles of average size 15-20 nm. HRTEM images reveal the preferential exposure of (101) crystallographic planes (see Fig. S10). Similar studies for CdS revealed presence of comparatively larger particles with average size of 40 nm. TEM studies for hexagonal AlN revealed formation of nanosheets with growth along (100) direction. Phase purity of all samples have also been confirmed by PL, UV, IR and Raman spectroscopic analyses (see Fig. S15-18), followed by surface topography analysis of sensors using 3D optical non-contact profilometer (see Fig. S19). Room temperature UV-Vis studies were carried out in a CECIL Aquarius 7200 spectrophotometer from 200 nm to 800 nm with scan rate of 10 nm/S (Fig. S15). Photoluminescence (PL) studies were carried out in a Fluoro Max-P (HORIBA JobinYvon) luminescence spectrophotometer (Fig. S16). In order to characterize the bonding patterns of samples, Fourier Transform Infrared (FTIR) studies were carried out in a Perkin Elmer FTIR instrument (Fig. S17). For sample preparation, small amount of sample was thoroughly mixed with dry analytical grade KBr powder, pressed into pellets and scanned from 470 to 4000 cm$^{-1}$. Samples were further analyzed by room temperature Raman studies using an excitation by 514 nm laser source in Renishaw inVia Microscope (Fig. S18). The phonon vibration modes were checked for ensuring sample purity.

**Calculations for ERSI and TPSA**

The dynamic response curve (resistance vs. time graph) for each gas/VOC sensing was used to obtain the values of $R_{air}$ (sensor resistance in absence of gas) and $R_{gas}$ (sensor resistance in presence of gas) along with the $\tau$ (response time, calculated in terms of time period required by the sensor to reach minimum/maximum resistance on application of target gas/VOC) (see Fig. S20-S25). Then using formula 4 in main text, the energy change involved is obtained as $\Delta E = K \times T \times \ln(\frac{R_{air}}{R_{gas}})$. ERSI is finally obtained by dividing $\Delta E$ with $\tau$ (see Table S2). TPSA values of different molecules have been calculated using Cactvs and OpenEye software based on the summation of tabulated surface contributions of polar fragments in the target molecule. The contributions of individual polar fragments were fitted by least squares method and compared with 3D PSA values. The TPSA concept is a modified version of calculating polar surface area of molecules where the 3D model of molecule was used in identifying the polar areas linked to a molecule and adding them up. TPSA concept was initially used in predicting transport properties and preferential adsorption of drug molecules through different types of body membranes with smaller TPSA molecules having greater chances of penetration. We have extended the concept for explaining greater surface adsorption by smaller TPSA molecules and the mutual interaction of their polar surface areas with the active adsorption sites generated on sensor surface. Both cases are perceived to be similar in behaviour; the competitive penetration of drug molecules through the cell membrane and cross-responses arising due to variable interactions of gas molecules with different TPSA on the sensor surface.

Table S1: Examples to illustrate the scheme for calculation of ERSI values.

| Material | Gas/VOC | $R_{air}$ (Ω) | $R_{gas}$ (Ω) | Response time (S) |
|---|---|---|---|---|
| $SnO_2$ | CO | $1 \times 10^3$ | $0.885 \times 10^3$ | 3 |
| $TiO_2$ | $H_2S$ | $50 \times 10^6$ | $45 \times 10^6$ | 5 |
| | | $8 \times 10^6$ | $6.88 \times 10^6$ | 3 |
| $Sn_{0.696}V_{0.304}O_2$ | CO | | | |
| | | $8 \times 10^6$ | $5.6 \times 10^6$ | 3 |
| | $H_2S$ | | | |
| | | $8 \times 10^6$ | $4.0 \times 10^6$ | 2 |
| | $NH_4Cl$ | | | |
| | $NH_3$ | $8 \times 10^6$ | $1.76 \times 10^6$ | 2 |

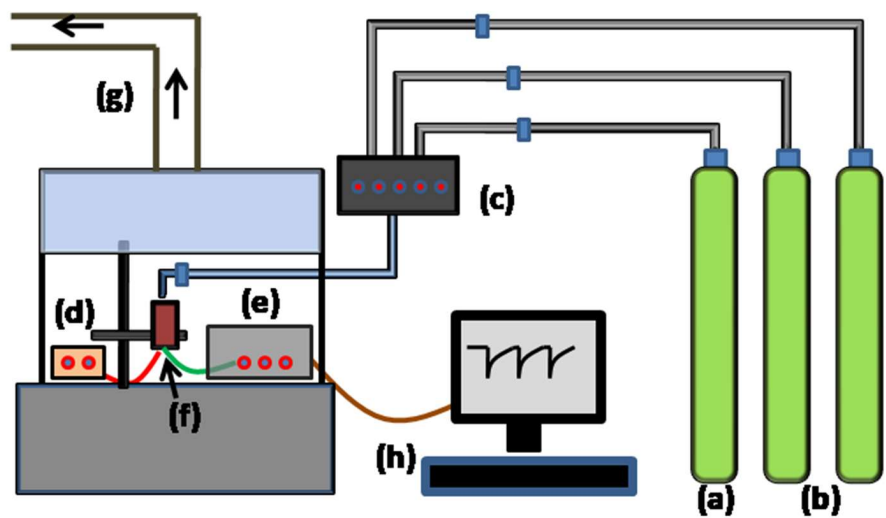

(a) Nitrogen cylinder (b) gas cylinder (c) mass flow controller with switches (d) DC power supply (e) Multimeter (f) sensor module (g) Fume hood (h) computer

Fig S1: A schematic diagram of the gas sensing experiments.

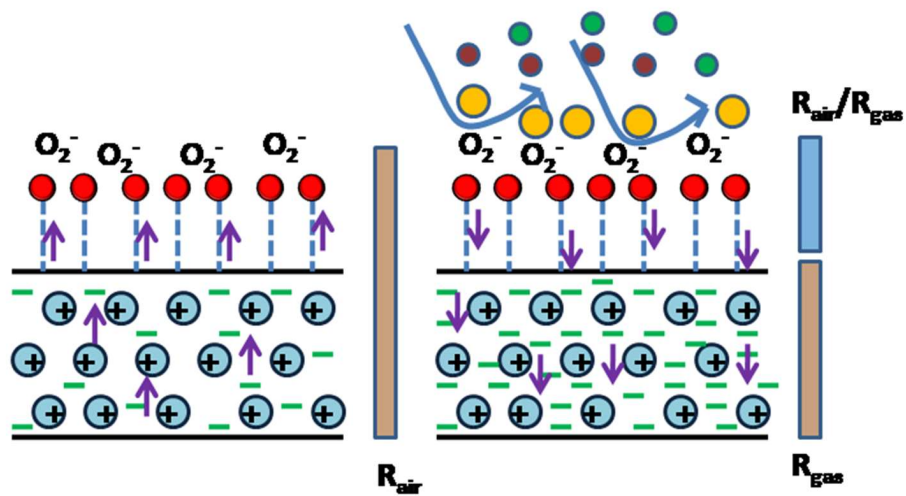

Fig S2: A schematic of surface interaction energy calculation. The bars in red colour highlight magnitude of sample resistance, which changes on account of surface gaseous interactions. The relative change in resistance as a result of surface reaction is marked by the blue bar.

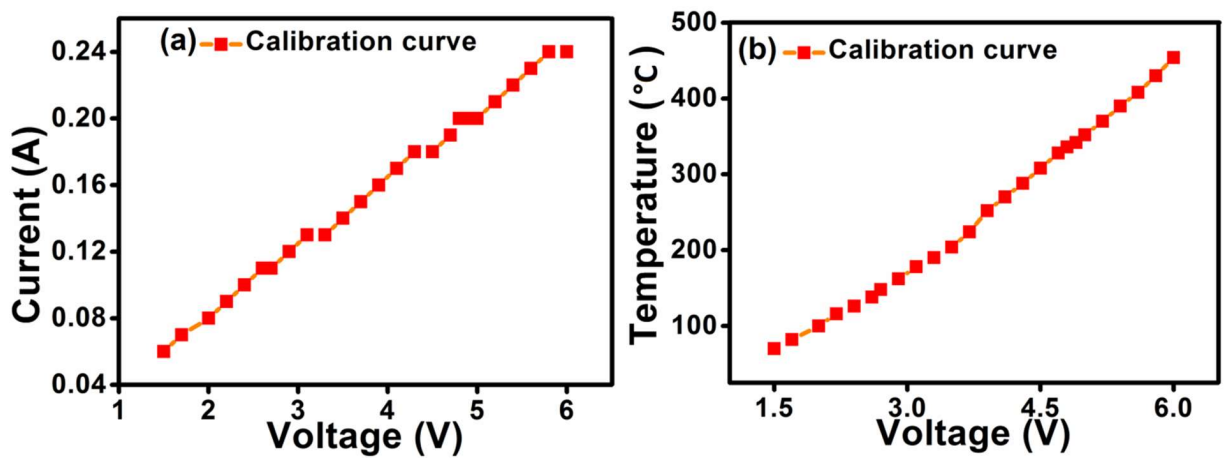

Fig S3: Calibration curves (a) Current vs. voltage and (b) Temperature vs. voltage for the sensor system(s) used.

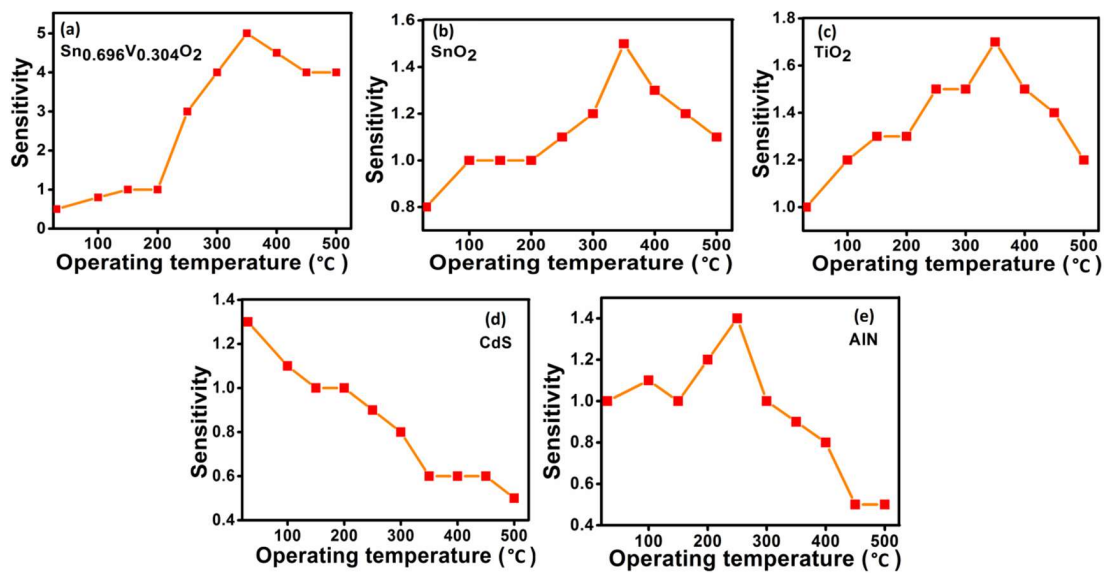

Fig S4: Sensitivity vs. operating temperature curve for (a) $Sn_{0.696}V_{0.304}O_2$ (b) $SnO_2$ (c) $TiO_2$ (anatase) (d) CdS (e) AlN sensors respectively. While for (a, b and c) the sensitivity was maximum at operating temperature of 350°C, that for (d) was at room temperature and for (e) was at 250°C. Dynamic resistance studies were carried out at the aforementioned temperatures for individual samples.

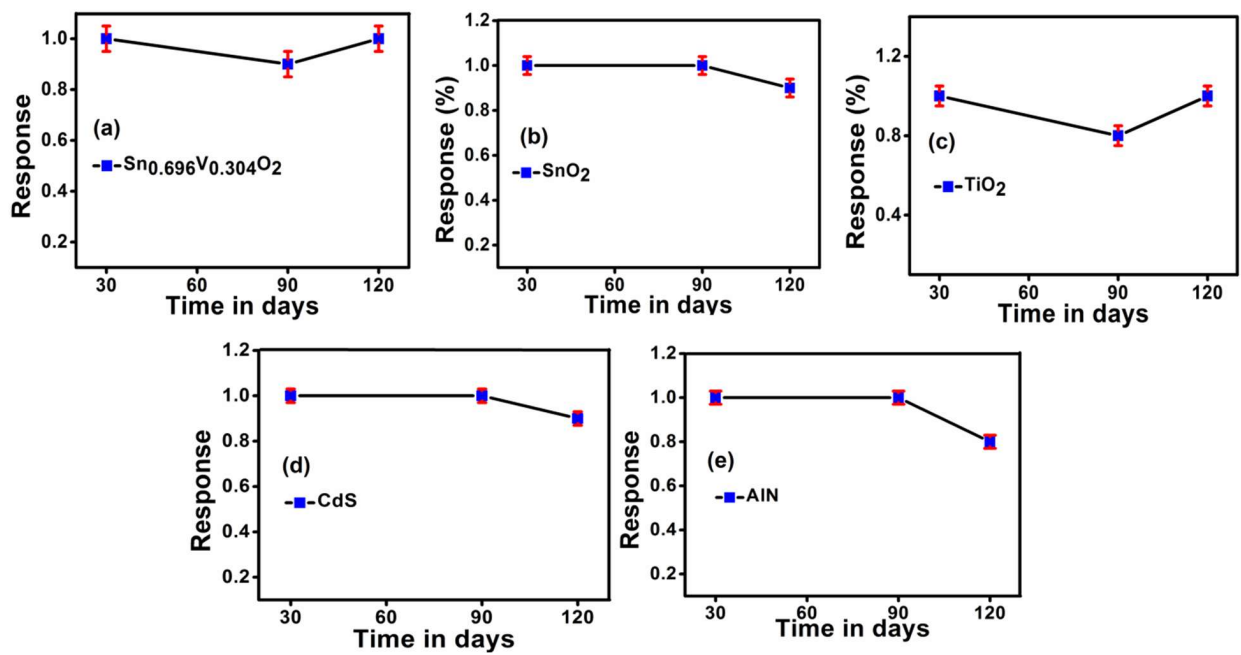

Fig S5: Stability analyses of (a) $Sn_{0.696}V_{0.304}O_2$ (b) $SnO_2$ (c) $TiO_2$ (anatase) (d) CdS (e) AlN sensors respectively over a time span of 4 months with response uncertainties expressed by error bars (in red). The low uncertainty in stability values re-confirm the long shelf life of the materials used and reproducibility of data obtained from them.

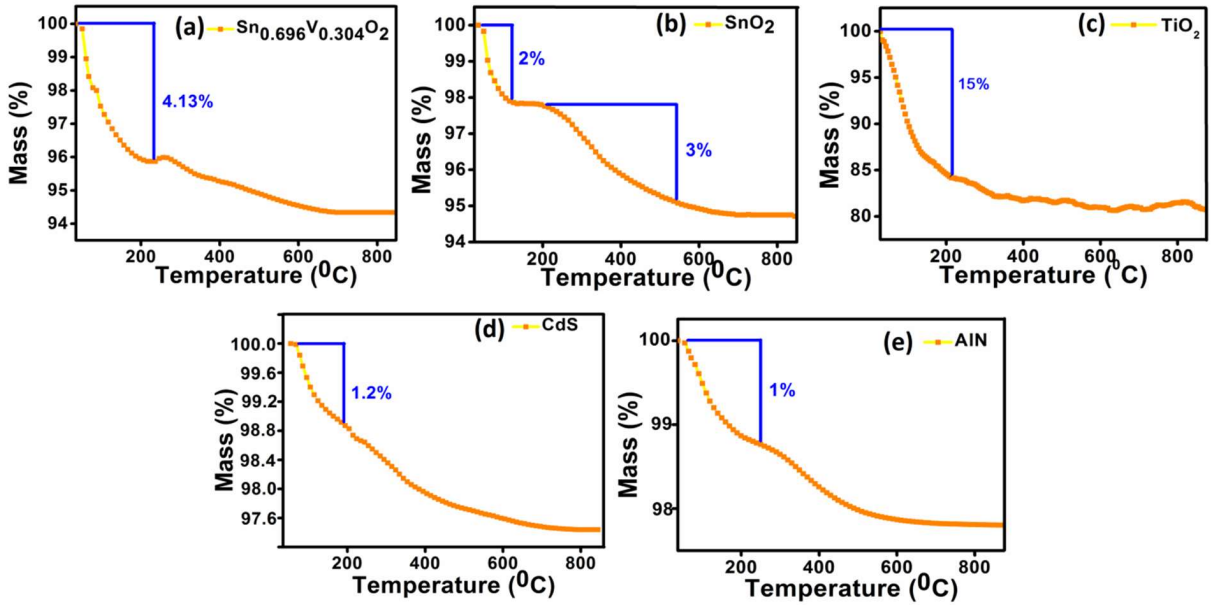

Fig S6: Thermo-Gravimetric Analysis of (a) $Sn_{0.696}V_{0.304}O_2$ (b) $SnO_2$ (c) $TiO_2$ (anatase) (d) CdS (e) AlN samples respectively, in air medium. Samples "a", "b" were further heat treated at 650°C and sample "c" at 450°C to obtain pure phase material. Samples "d" and "e" were already procured in pure phase conditions. Negligible mass loss of around 1% owing to removal of surface adsorbed moisture indicates absence of any impurity/spurious phase.

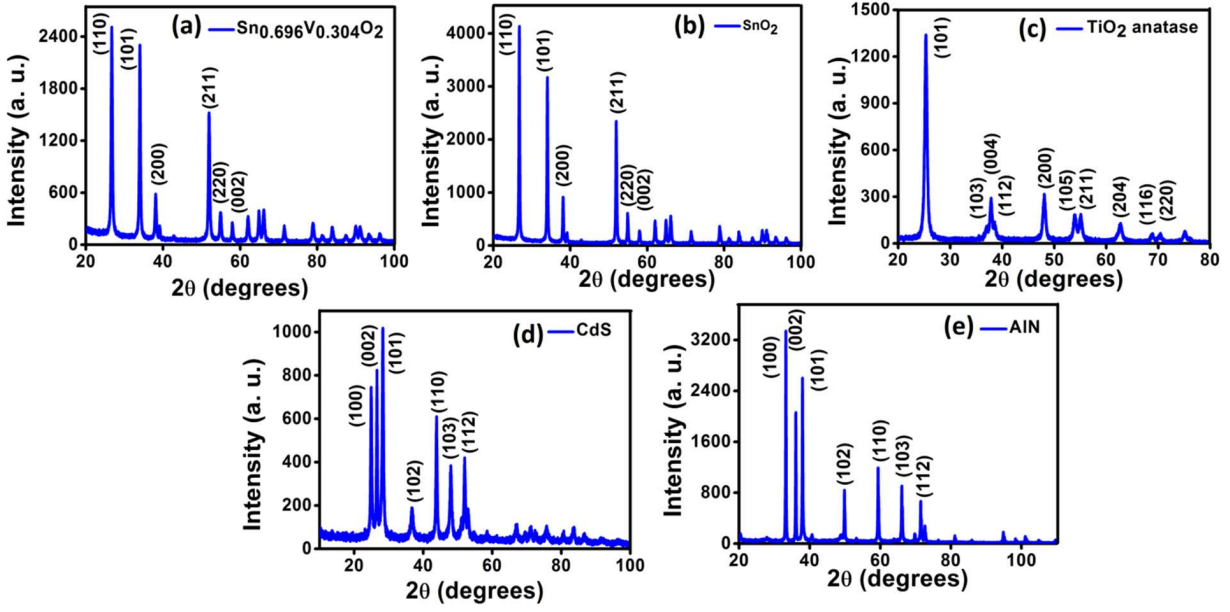

Fig S7: Room temperature X-ray diffraction pattern(s) of (a) $Sn_{0.696}V_{0.304}O_2$ (b) $SnO_2$ (c) $TiO_2$ (anatase) (d) CdS (e) AlN samples respectively. The (h k l) notations correspond to a particular set of planes by Miller indexing. ($Sn_{0.696}V_{0.304}O_2$ and $SnO_2$ are in tetragonal rutile phase, $TiO_2$ is in anatase phase, CdS is in hexagonal Wurtzite phase and AlN is in hexagonal phase)

Table S2: Structural information of all the samples with $a$, $b$, $c$ in Å.

| Sample | Space group | Lattice parameters |
|---|---|---|
| $Sn_{0.696}V_{0.304}O_2$ (rutile) | $P4_2/mnm$ | $a$=4.7312(1), $c$=3.1824(3) $\alpha=\beta=\gamma=90°$ |
| $SnO_2$ (rutile) | $P4_2/mnm$ | $a$=4.7345(8), $c$=9.5053(6) $\alpha=\beta=\gamma=90°$ |
| $TiO_2$ (anatase) | I 41/a m d | $a$=3.7843(2), $c$=9.5053(6) $\alpha=\beta=\gamma=90°$ |
| CdS (hexagonal) | P63/mc | $a$=4.1332(2), $c$=6.7147(5) $\alpha=\beta=90°, \gamma=120°$ |
| AlN (hexagonal) | P63/mc | $a$=3.1112(5), $c$=4.9793(7) $\alpha=\beta=90°, \gamma=120°$ |

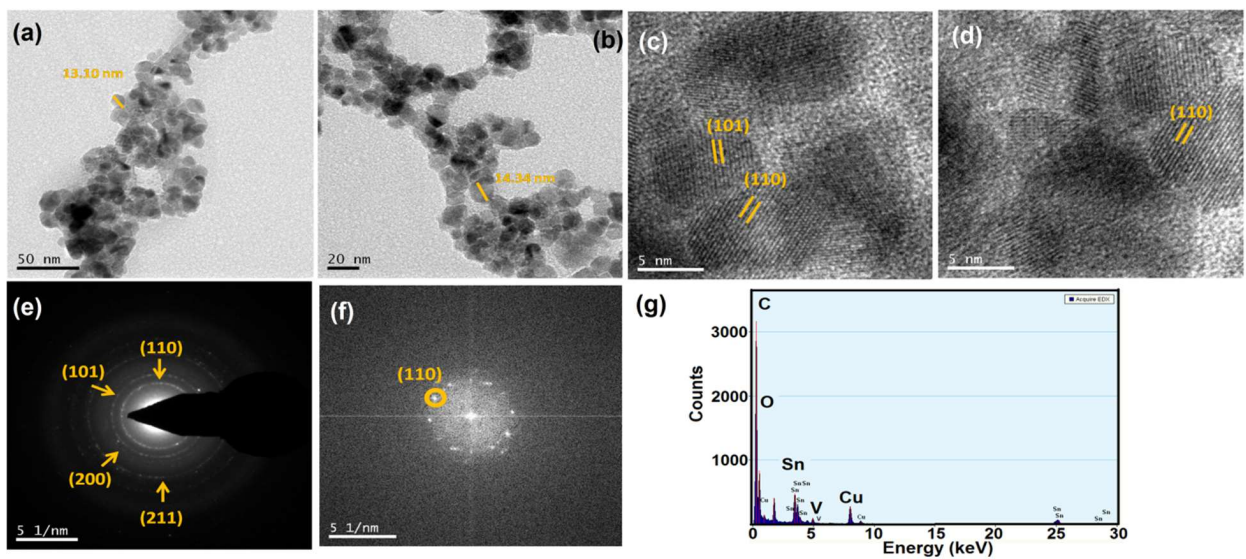

Fig S8: TEM results of $Sn_{0.696}V_{0.304}O_2$ sample. (a, b) bright field images showing nano-sized particle formation with spherical morphology in range of 13-15 nm (c, d) HRTEM images highlighting preferential exposure of (110) and (101) crystallographic planes (e) SAED pattern showing the Bragg circles (f) reduced FFT pattern of (110) planes in fig "d" (g) EDX spectrum showing the major elements Sn, V and O. The signals of Cu and C are due to use of copper coated carbon grids during microscopic studies.

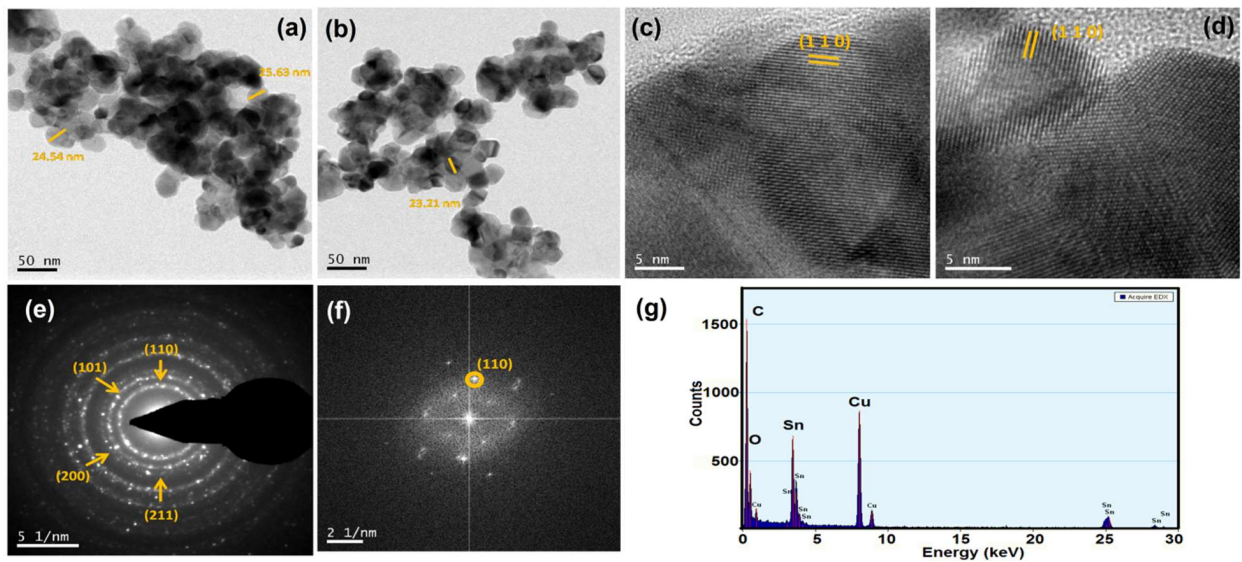

Fig S9: TEM results of SnO$_2$ sample. (a, b) bright field images showing nano-sized particle formation with spherical morphology in range of 14-15 nm (c, d) HRTEM images highlighting preferential exposure of (110) and (101) crystallographic planes (e) SAED pattern showing the Bragg circles (f) reduced FFT pattern of (110) planes in fig "d" (g) EDX spectrum showing the major elements Sn and O. The signals of Cu and C are due to use of copper coated carbon grids during microscopic studies.

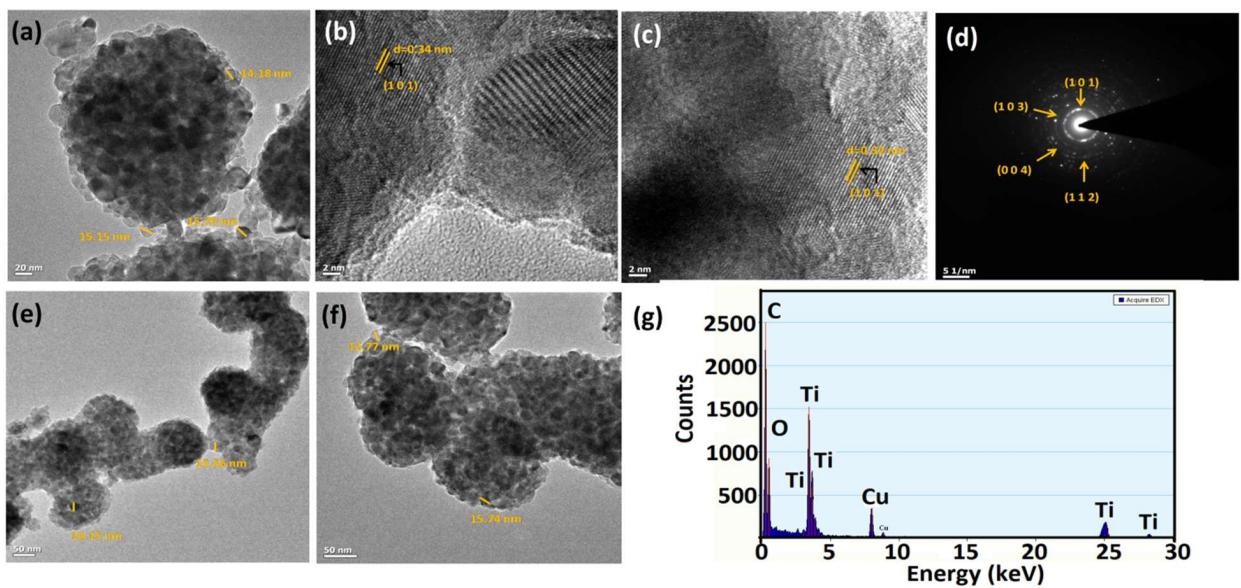

Fig S10: TEM results of TiO$_2$ sample. (a, e and f) bright field images showing nano-sized particle formation with spherical morphology in range of 13-15 nm (b, c) HRTEM images highlighting preferential exposure of (101) crystallographic planes (d) SAED pattern showing the Bragg circles (g) EDX spectrum showing the major elements Ti and O. The signals of Cu and C are due to use of copper coated carbon grids during microscopic studies.

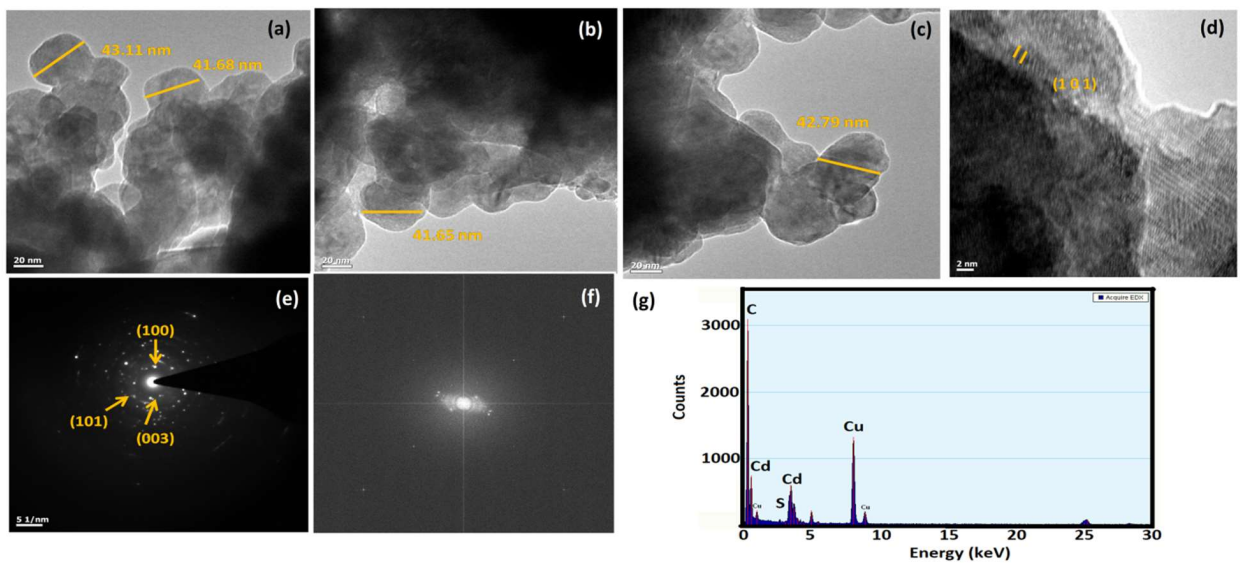

Fig S11: TEM results of CdS sample. (a, b and c) bright field images showing particle formation with spherical morphology in range of 40-45 nm (d) HRTEM image of CdS powder with preferential exposure of (101) planes (e) SAED pattern showing the Bragg rings (f) FFT images of (101) planes (g) EDX spectrum of CdS showing the major elements Cd and S.

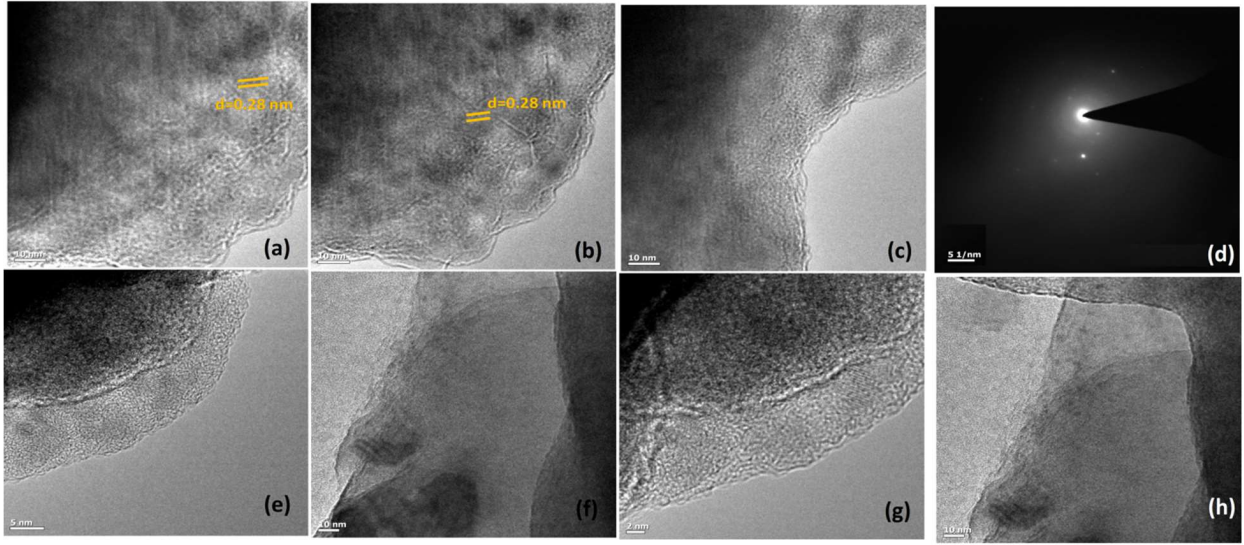

Fig S12: TEM results of AlN sample. (a and b) HRTEM micrographs revealing growth of AlN sheets in (100) direction, indicated by d spacing of 0.28 nm(c, e, f, g and h) bright field images revealing existence of sheet like structures/morphology (d) SAED pattern.

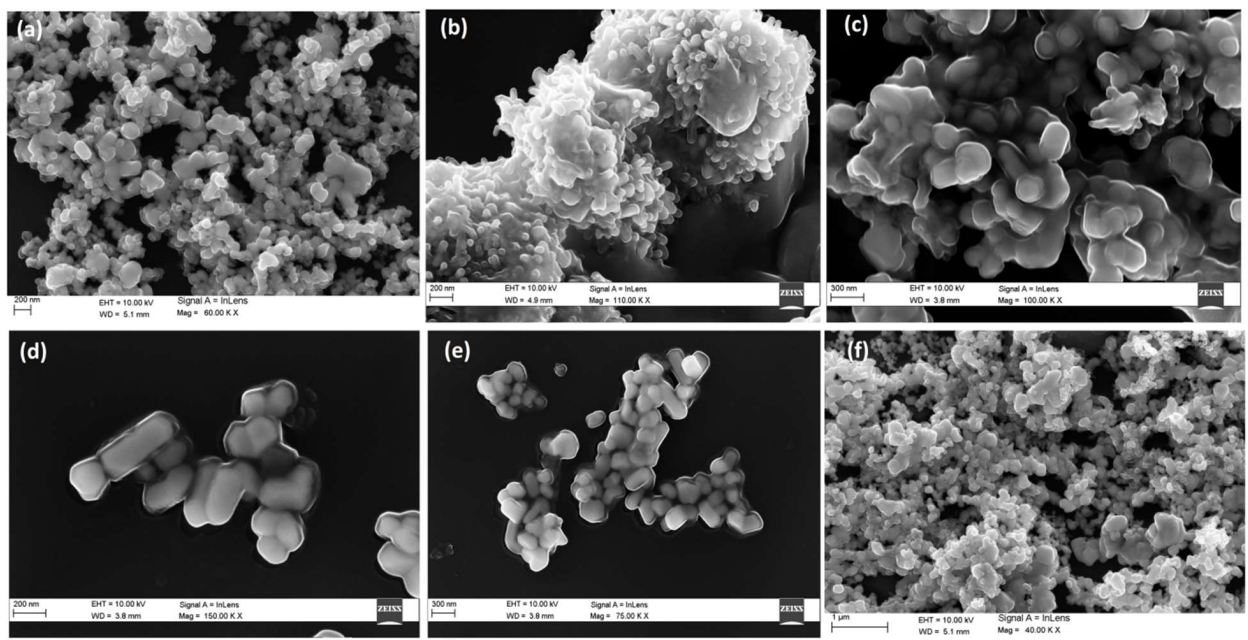

Fig S13: FESEM images of (a) $SnO_2$ (b, c d and e) $Sn_{0.696}V_{0.304}O_2$ (f) $TiO_2$ samples respectively. Formation of spherical nanoparticles is visible from the micrographs.

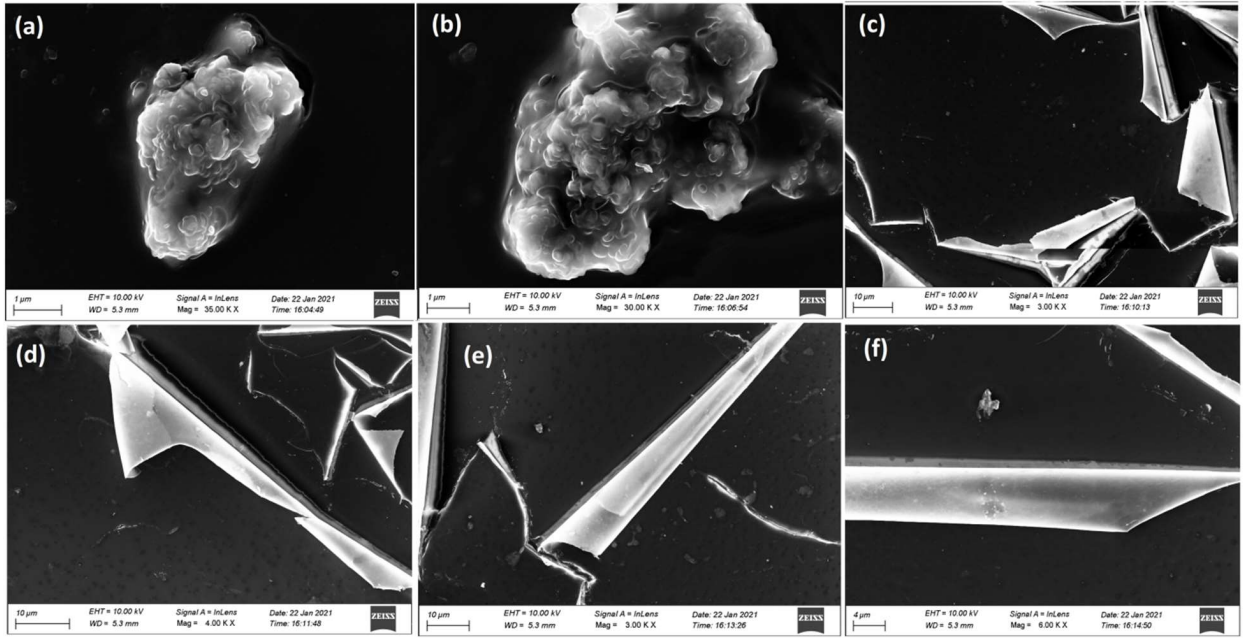

Fig S14: FESEM images of (a and b) CdS (c, d, e and f) AlN samples respectively. While CdS sample exist with spherical morphology, AlN is present with a 2D sheet like appearance which is in consonance with TEM results as well.

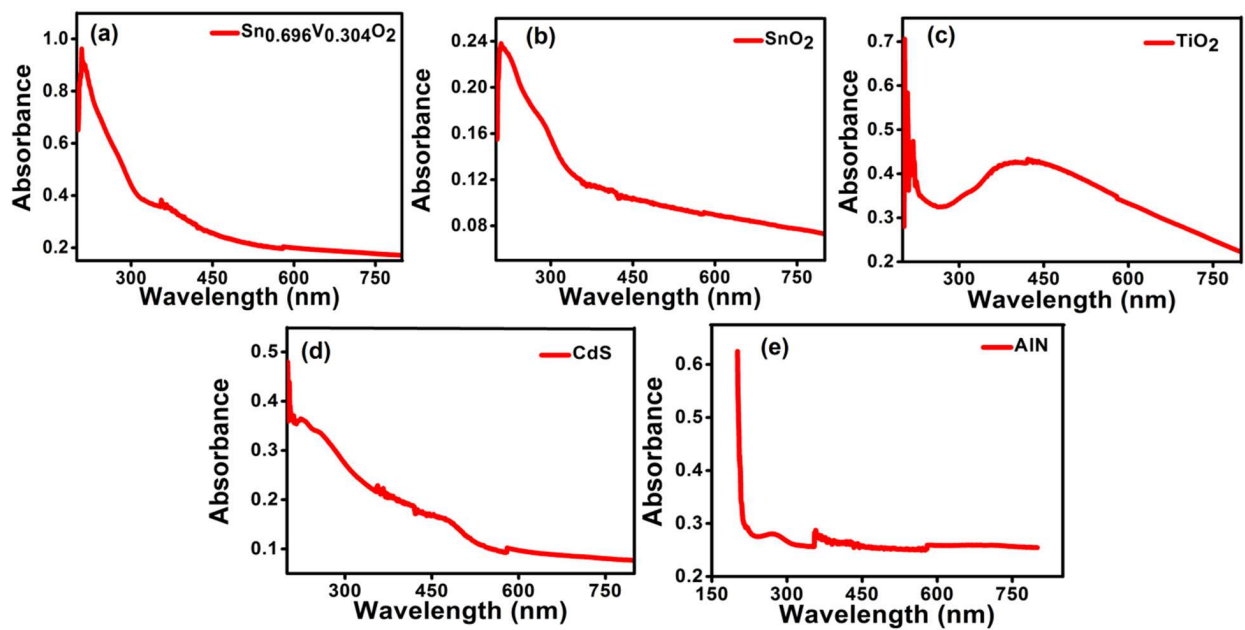

Fig S15: Room temperature UV-Vis spectra of (a) $Sn_{0.696}V_{0.304}O_2$ (b) $SnO_2$ (c) $TiO_2$ (anatase) (d) CdS (e) AlN samples respectively.

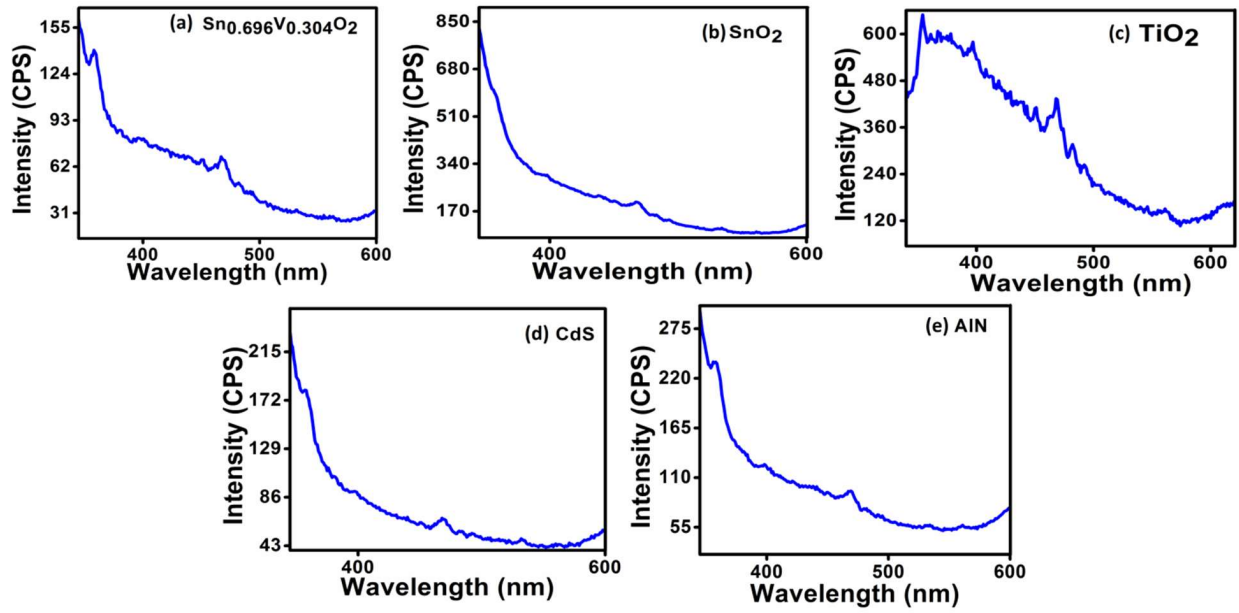

Fig S16: Room temperature PL studies of (a) $Sn_{0.696}V_{0.304}O_2$ (b) $SnO_2$ (c) $TiO_2$ (anatase) (d) CdS (e) AlN samples respectively under excitation by 325 nm.

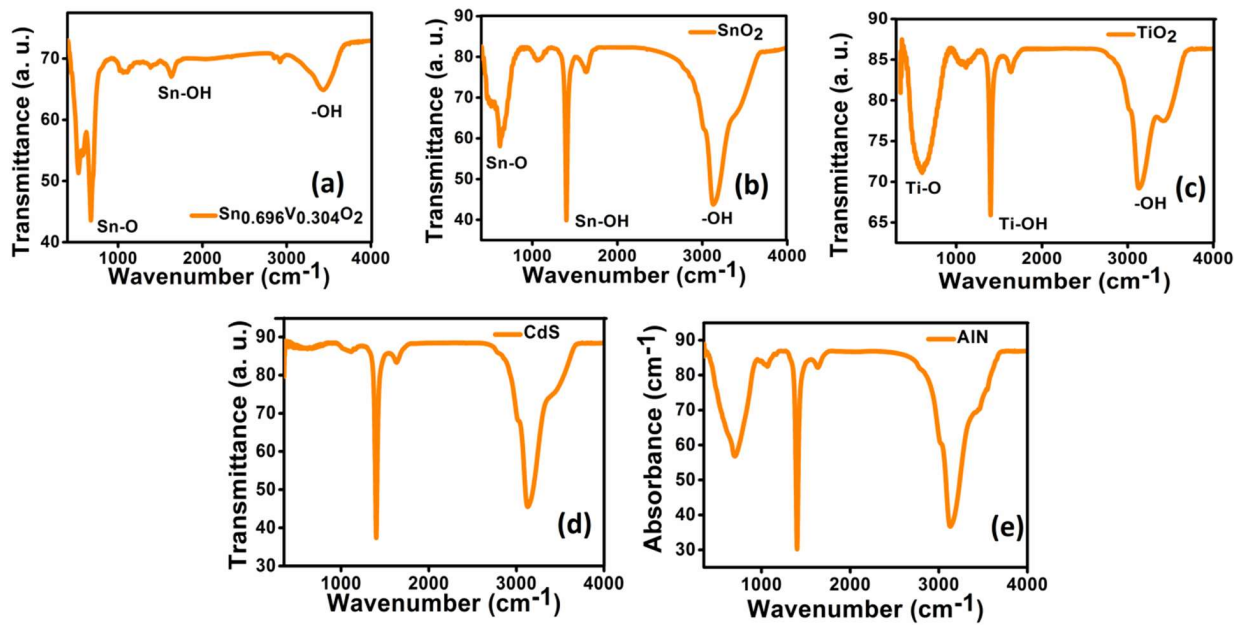

Fig S17: Room temperature FTIR spectra of (a) $Sn_{0.696}V_{0.304}O_2$ (b) $SnO_2$ (c) $TiO_2$ (anatase) (d) CdS (e) AlN samples respectively. The –OH vibration signals are due to presence of surface moisture. The spectra indicate formation of pure phase materials.

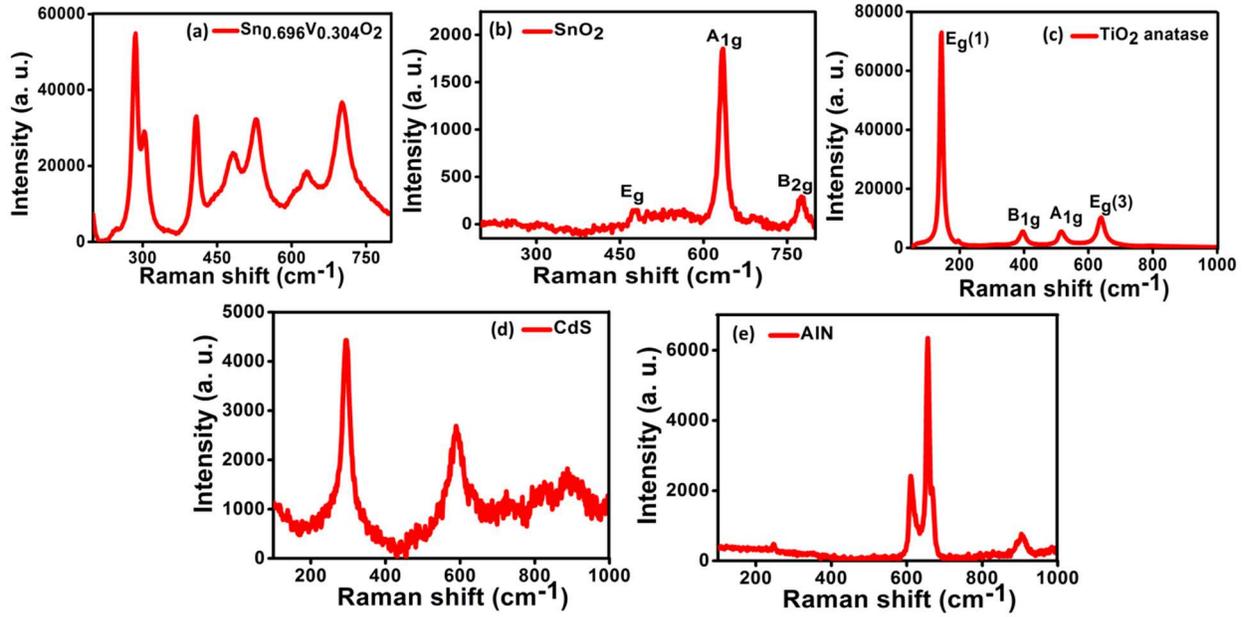

Fig S18: Room temperature Raman spectra of (a) $Sn_{0.696}V_{0.304}O_2$ (b) $SnO_2$ (c) $TiO_2$ (anatase) (d) CdS (e) AlN samples respectively under excitation by laser source of 514 nm.

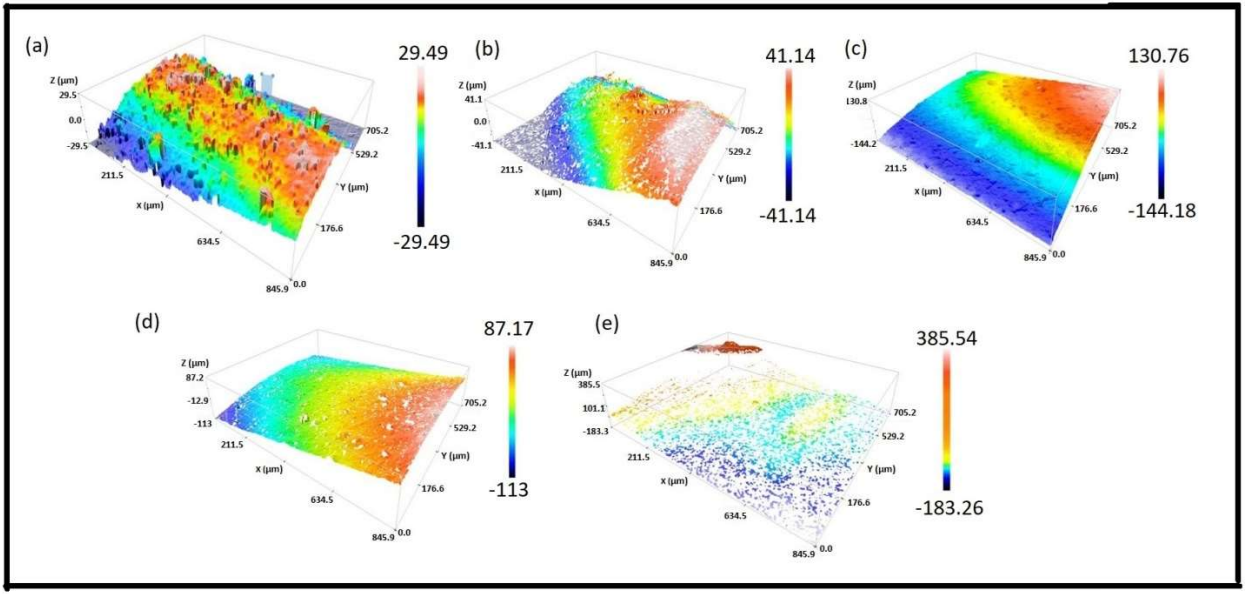

Fig. S19: Surface topography analysis of thick film coatings. Surface topography of (a) CdS (b) $SnO_2$ (c) $Sn_{0.696}V_{0.304}O_2$ (d) $TiO_2$ (anatase) (e) AlN coatings respectively with gradient scale in µm. All images were taken using non-contact 3D optical profiler in focus variation mode.

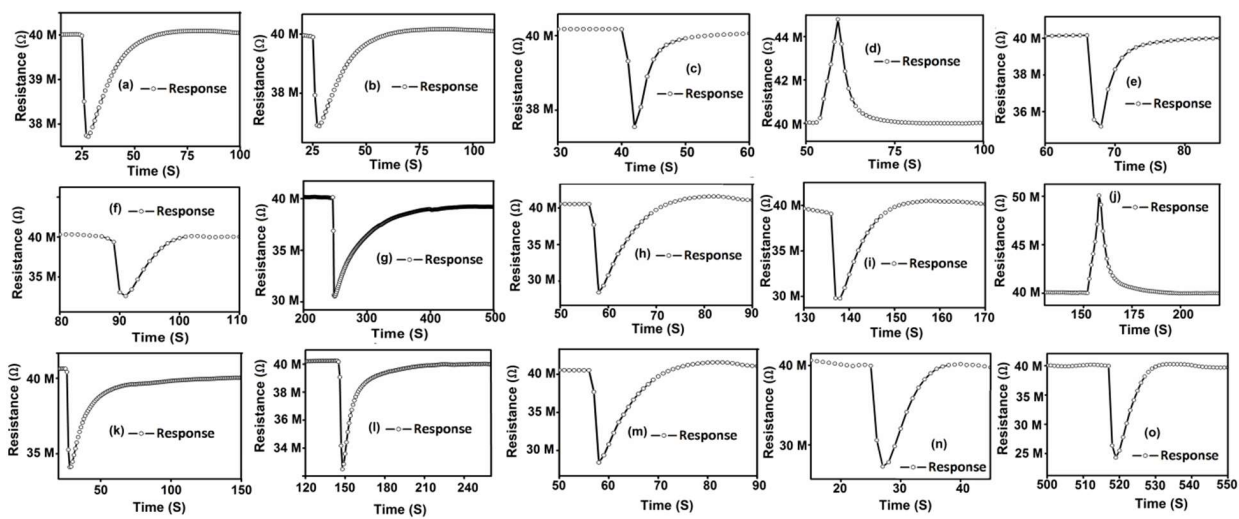

Fig. S20: Dynamic resistance measurement for CdS. Dynamic resistance curve of CdS on interaction with 10 ppm of (a) benzophenone (b) CO (c) $H_2S$ (d) $SO_2$ (e) $NH_4Cl$ (f) methanol (g) $NH_3$ (h) iso-amyl alcohol (i) benzaldehyde (j) NO (k) formaldehyde (l) pyran (m) iso-propyl alcohol (n) ethanol (o) acetone respectively.

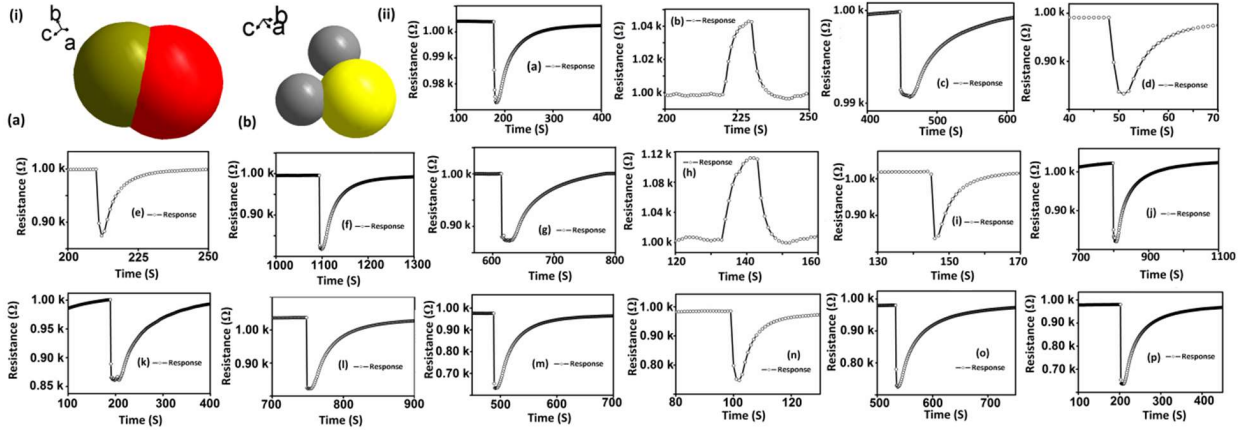

Fig. S21: Dynamic resistance measurement for SnO$_2$.(i) Molecular representation of (a) CO (green: C, red: O) b) H$_2$S (grey: H, yellow: S) using ChemDraw 3D software.(ii)Dynamic resistance curve of SnO$_2$ on interaction with 10 ppm of (a) ethylene glycol (b) SO$_2$ (c) benzophenone (d) methanol (e) formaldehyde (f) CO (g) iso-amyl alcohol (h) NO (i) benzaldehyde (j) H$_2$S (k) pyran (l) iso-propyl alcohol (m) ethanol (n) acetone (o) NH$_4$Cl (p)NH$_3$ respectively.

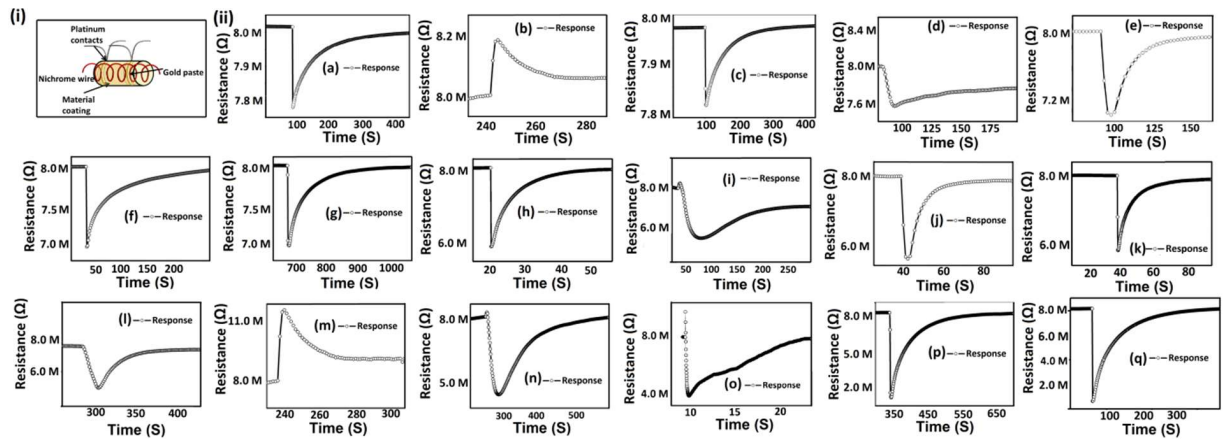

Fig. S22: Dynamic resistance measurement for Sn$_{0.696}$V$_{0.304}$O$_2$.(i) A schematic representation of sensor module (ii)Dynamic resistance curve of Sn$_{0.696}$V$_{0.304}$O$_2$ on interaction with 10 ppm of (a) ethylene glycol (b) SO$_2$ (c) aniline (d) methanol (e) benzophenone (f) formaldehyde (g) CO (h) iso-amyl alcohol (i) H$_2$S(j) benzaldehyde (k) iso-propyl alcohol (l) ethanol (m) NO (n) acetone (o) NH$_4$Cl (p) NH$_3$ (q) tetrahydrofuran respectively.

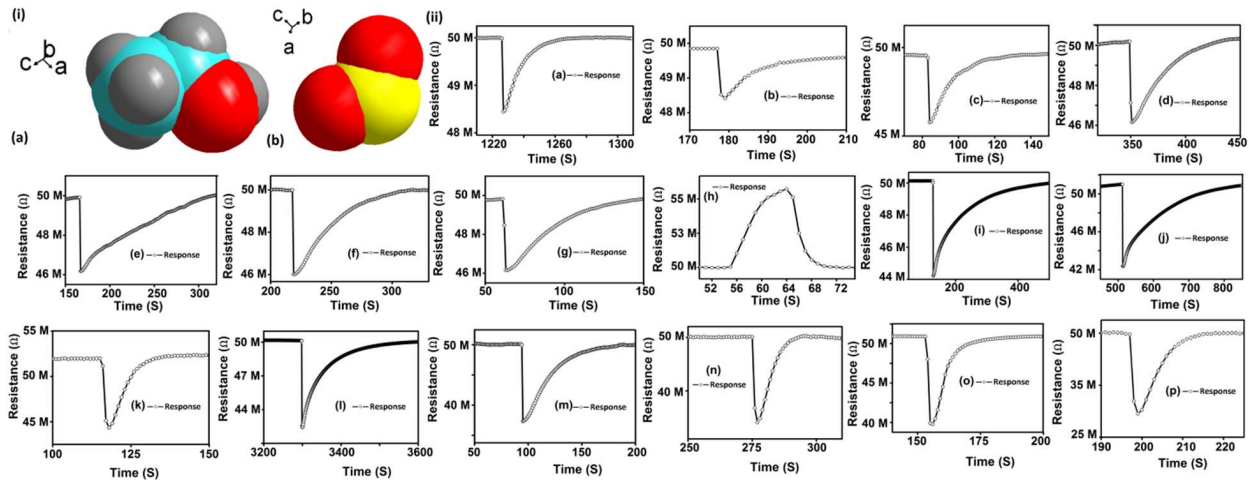

Fig. S23: Dynamic resistance measurement for TiO$_2$. (i) Molecular representation of (a) ethanol (blue: C, grey: H, red: O) (b) SO$_2$ (yellow: S, red: O) using ChemDraw 3D software. (ii) Dynamic resistance curve of TiO$_2$ on interaction with 10 ppm of (a) ethylene glycol (b) aniline (c) benzophenone (d) benzaldehyde (e) methanol (f) iso-amyl alcohol (g) H$_2$S (h) NO (i) formaldehyde (j) CO (k) pyran (l) NH$_4$Cl (m) iso-propyl alcohol (n) ethanol (o) acetone (p) NH$_3$ respectively.

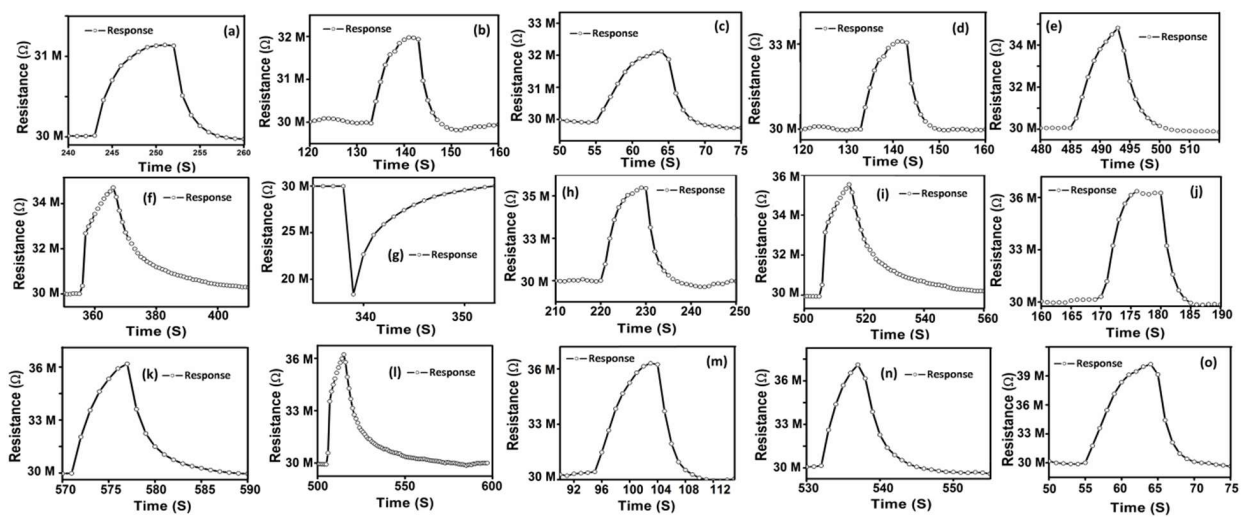

Fig. S24: Dynamic resistance measurement for AlN. Dynamic resistance curve of AlN on interaction with 10 ppm (a) benzophenone (b) benzaldehyde (c) formaldehyde (d) ethylene glycol (e) CO (f) methanol (g) NO (h) acetone (i) H$_2$S (j) pyran (k) NH$_4$Cl (l) iso-propyl alcohol (m) NH$_3$ (n) iso-amyl alcohol (o) ethanol respectively.

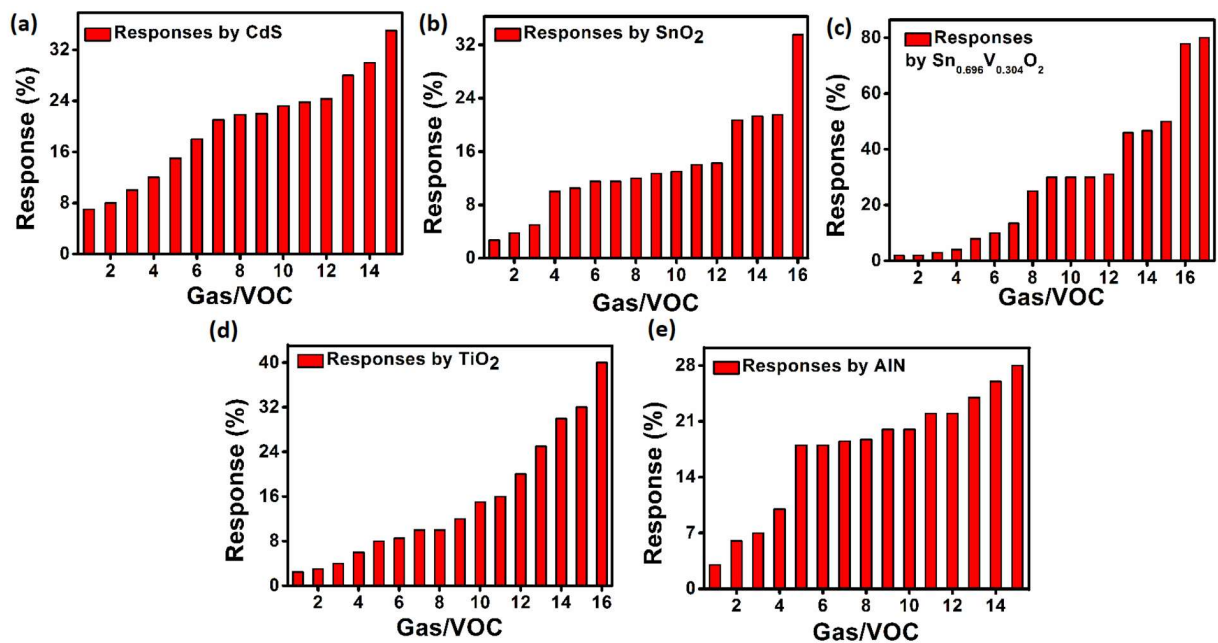

Fig. S25: Cross response. Cross response plots for (a) CdS on interaction with 10 ppm (1) benzophenone (2) CO (3) $H_2S$ (4)$SO_2$ (5) $NH_4Cl$ (6) methanol (7) $NH_3$ (8) iso-amyl alcohol (9) benzaldehyde (10) NO (11) formaldehyde (12) pyran (13) iso-propyl alcohol (14) ethanol (15) acetone respectively (b) $SnO_2$ on interaction with 10 ppm (1) ethylene glycol (2) $SO_2$ (3) benzophenone (4) methanol (5) formaldehyde (6) CO (7) iso-amyl alcohol (8) NO (9) benzaldehyde (10) $H_2S$ (11) pyran (12) iso-propyl alcohol (13) ethanol (14) acetone (15) $NH_4Cl$ (16)$NH_3$ respectively (c) $Sn_{0.696}V_{0.304}O_2$ on interaction with 10 ppm (1) ethylene glycol (2) $SO_2$ (3) aniline (4) methanol (5) benzophenone (6) formaldehyde (7) CO (8) iso-amyl alcohol (9) $H_2S$(10) benzaldehyde (11) iso-propyl alcohol (12) ethanol (13) NO (14) acetone (15) $NH_4Cl$ (16) $NH_3$ (17) tetrahydrofuran respectively(d) $TiO_2$ on interaction with 10 ppm (1) ethylene glycol (2) aniline (3) benzophenone (4) benzaldehyde (5) methanol (6) iso-amyl alcohol (7) $H_2S$ (8) NO (9) formaldehyde (10) CO (11) pyran (12) $NH_4Cl$ (13) iso-propyl alcohol (14) ethanol (15) acetone (16) $NH_3$ respectively (e) AlN on interaction with 10 ppm (1) benzophenone (2) benzaldehyde (3) formaldehyde (4) ethylene glycol (5) CO (6) methanol (7) NO (8) acetone (9) $H_2S$ (10) pyran (11) $NH_4Cl$ (12) iso-propyl alcohol (13) $NH_3$ (14) iso-amyl alcohol (15) ethanol respectively.

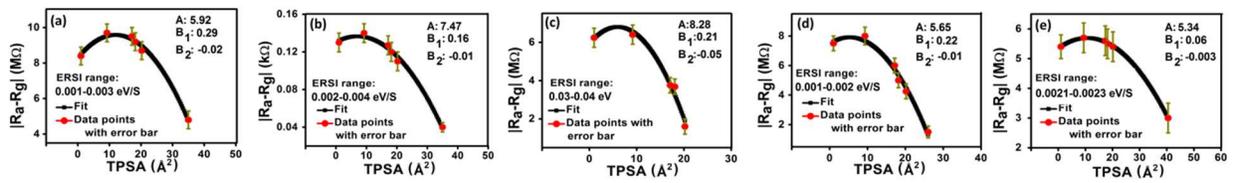

Fig. S26. Quadratic (y=Ax+$B_1$x+$B_2$$x^2$, y=response, x=TPSA) correlation between surface interaction and TPSA.(i) Change in resistance vs. TPSA plots for (a) CdS (b) $SnO_2$ (c) $Sn_{0.696}V_{0.304}O_2$ (d) $TiO_2$ (anatase) (e) AlN samples respectively.

Table S3: ERSI and response data. ERSI values (absolute) for surface interactions with gas/VOC molecules of different TPSA values along with the operational temperature in brackets.

| TPSA ($Å^2$) | Gas/VOC | ERSI (eV/S)(x$10^{-3}$) | | | | | | | | |
|---|---|---|---|---|---|---|---|---|---|---|
| | | $Sn_{0.696}V_{0.304}O_2$ (350°C) | | $SnO_2$ (350°C) | | $TiO_2$ (350°C) | | CdS (25°C) | | AlN (250°C) | |
| 1 | CO | 2.6 | 13.5 | 2.1 | 11.5 | 2.1 | 15 | 0.53 | 8 | 2.22 | 18 |
| | $H_2S$ | 6.5 | 30 | 3.7 | 13 | 1.1 | 10 | 0.90 | 10 | 2.44 | 20 |
| | $NH_4Cl$ | 20.1 | 50 | 4.3 | 21.5 | 3.9 | 20 | 1.39 | 15 | 2.67 | 22 |
| | $NH_3$ | 49.8 | 77.7 | 10.9 | 33.5 | 9.1 | 40 | 2.86 | 21 | 3.85 | 24 |
| 9.2 | Pyran/THF | 28.8 | 80 | 2.70 | 14 | 1.8 | 16 | 2.385 | 24.3 | 2.17 | 20 |
| 17.1 | Benzophenone | 1.1 | 30 | 0.65 | 5 | 0.052 | 4 | 0.932 | 7 | 0.794 | 3 |
| | Formaldehyde | 1.2 | 10 | 1.98 | 10.5 | 2.1 | 12 | 2.464 | 23.8 | 1.82 | 7 |
| | Benzaldehyde | 6.1 | 30 | 2.42 | 12.7 | 0.062 | 6 | 2.128 | 22 | 1.57 | 6 |
| | Acetone | 34 | 46.6 | 6.42 | 21.3 | 10.3 | 32 | 5.550 | 35 | 2.302 | 18.7 |
| 18.1 | NO | 41.2 | 46 | 2.18 | 12 | 1.8 | 10 | 1.77 | 23.2 | 2.176 | 18.5 |
| 20.2 | Methanol | 0.21 | 4 | 1.86 | 10 | 1.0 | 8 | 1.7 | 18 | 2.223 | 18 |
| | Iso-amyl alcohol | 5 | 25 | 2.17 | 11.5 | 1.3 | 8.5 | 2.86 | 21.8 | 4.1 | 26 |
| | Iso-propyl alcohol | 33.9 | 30 | 2.75 | 14.3 | 5.1 | 25 | 4.22 | 28 | 2.67 | 22 |
| | Ethanol | 26.2 | 31.1 | 6.20 | 20.7 | 9.5 | 30 | 4.58 | 30 | 4.4 | 28 |
| 26 | Aniline | 0.020 | 3 | — | — | 1.6 | 3 | — | — | — | — |
| 35 | $SO_2$ | 0.009 | 2 | 2.09 | 3.8 | — | — | 1.94 | 12 | — | — |
| 40.5 | Ethylene glycol | 0.006 | 2 | 0.29 | 2.7 | 0.27 | 2.5 | — | — | 2.048 | 10 |

Table S4: Fit values of response vs. ERSI curve for fixed TPSA. Power law fitting values ($y=Ax^b$) for response vs. ERSI curve. The quantity in bracket indicates the maximum uncertainty in second decimal place of the power term.

| TPSA ($Å^2$) | $Sn_{0.696}V_{0.304}O_2$ (350°C) | | $SnO_2$ (350°C) | | $TiO_2$ (350°C) | | CdS (25°C) | | AlN (250°C) | |
|---|---|---|---|---|---|---|---|---|---|---|
| | A | b | A | b | A | b | A | b | A | b |
| 1 | 368.44 | 0.51(1) | 681.80 | 0.52(2) | 1043.22 | 0.53(1) | 647.12 | 0.58(5) | 298.09 | 0.45 (5) |
| 17.1 | 213.67 | 0.50(2) | 404.14 | 0.55(2) | 561.02 | 0.59(4) | 636.53 | 0.55(9) | 407.06 | 0.55 (4) |
| 20.2 | 78.37 | 0.46(1) | 327.54 | 0.48(3) | 461.17 | 0.47(3) | 526.19 | 0.53(6) | 595.07 | 0.56 (8) |

Table S5: Fit values of response & change in resistance vs. TPSA curve for fixed ERSI range. Polynomial fitting values for response and change in resistance vs. TPSA curve.

| ERSI (eV/S) range | Values of coefficients for polynomial $y=Ax+B_1x+B_2x^2$ (y: response, x: TPSA) | | | Values of coefficients for polynomial $y=Ax+B_1x+B_2x^2$ (y: change in resistance, x: TPSA) | | |
|---|---|---|---|---|---|---|
| | A | $B_1$ | $B_2$ | A | $B_1$ | $B_2$ |
| 0.001 | 14.95 | 0.32 | -0.03 | 5.65 | 0.22 | -0.01 |
| 0.002 | 20.69 | 0.55 | -0.02 | 5.92 | 0.29 | -0.02 |
| 0.002 | 12.94 | 0.20 | -0.01 | 7.47 | 0.16 | -0.01 |
| 0.01 | 76.06 | 1.89 | -0.20 | 8.28 | 0.21 | -0.05 |
| 0.0002 | 18.12 | 0.21 | -0.01 | 5.34 | 0.06 | -0.003 |